\pdfoutput=1
\documentclass[conference]{IEEEtran}

\IEEEoverridecommandlockouts

\usepackage{cite}
\usepackage{amsmath,amssymb,amsfonts}
\usepackage{algorithm}
\usepackage{algorithmic}
\usepackage{makecell}
\usepackage{graphicx}
\usepackage{textcomp}
\usepackage{xcolor}
\usepackage{kotex}
\usepackage{comment}
\usepackage{multirow}
\usepackage{array}
\usepackage{enumitem}

\newcolumntype{C}[1]{>{\centering\arraybackslash}m{#1}}

\newtheorem{remark}{Remark}

\def\BibTeX{{\rm B\kern-.05em{\sc i\kern-.025em b}\kern-.08em
    T\kern-.1667em\lower.7ex\hbox{E}\kern-.125emX}}

\begin{document}

\title{\LARGE Semantics-Aware Hierarchical Token Communication: \\ Clustering, Bit Mapping, and Power Allocation\vspace{-10pt}}
\author{
Jihoon Lee$^{*}$, Seungeun Oh$^{\dagger}$, Jihong Park$^{\dagger}$, Seong-Lyun Kim$^{*}$, and Seung-Woo Ko$^{\ddagger}$\\
$^{*}$Yonsei University, Korea, e-mail: \{jhlee, slkim\}@ramo.yonsei.ac.kr\\
$^{\dagger}$Singapore University of Technology and Design, Singapore, e-mail: \{seungeun\_oh, jihong\_park\}@sutd.edu.sg\\
$^{\ddagger}$Inha University, Korea, e-mail: swko@inha.ac.kr \vspace{-20pt}
} 

\maketitle
 
\begin{abstract}
Despite the rise of \emph{token communication} (TokCom) as a new paradigm beyond traditional bit communication, existing approaches have primarily adopted \emph{artificial intelligence} (AI)-centric designs that rely on semantic recovery via large models. Meanwhile, their physical-layer designs, such as token-bit mapping and power allocation, remain conventional and do not reflect token-level semantics. These semantics-agnostic designs can lead to significant semantic loss, particularly at low \emph{signal-to-noise ratio} (SNR) levels. To address this issue, we propose \emph{hierarchical TokCom} (H-TokCom), a framework that embeds semantic structure directly into physical-layer design. The key idea is to group semantically similar tokens into clusters and hierarchically assign their bit representations, where each token is represented by a cluster-level prefix and a token-specific suffix. As long as the cluster bits are correctly delivered, errors in the suffix bits typically map the received token to another within the same semantic cluster, resulting in only limited semantic distortion. This robustness is further strengthened by allocating more transmit power to the prefix bits than to the suffix bits. Simulation results show that H-TokCom achieves substantial semantic-similarity gains over conventional TokCom across the considered SNR range, increasing the semantic similarity from $0.206$ to $0.279$ at $\gamma=3$ dB on COCO, corresponding to a gain of $0.073$ $(35.4\%)$.
\end{abstract}

\vspace{-10pt}
\section{Introduction}

A token represents the fundamental processing unit of input and output sequences in a transformer, which underpins modern large foundation models \cite{GPT }. 
 The rapid evolution of these models, from standalone architectures to large-scale agentic ecosystems, has elevated the token into a universal information unit that extends far beyond the classic notion of a bit. The universality spans artificial intelligence, autonomous systems, and increasingly, communication systems. Motivated by this paradigm shift, our work builds upon the growing research direction that unifies AI and communication through a token-centric perspective, commonly referred to as \emph{token communication} (TokCom) \cite{qiao2025todma}.

\begin{figure}[t]
  \centering
  \includegraphics[width=\linewidth]{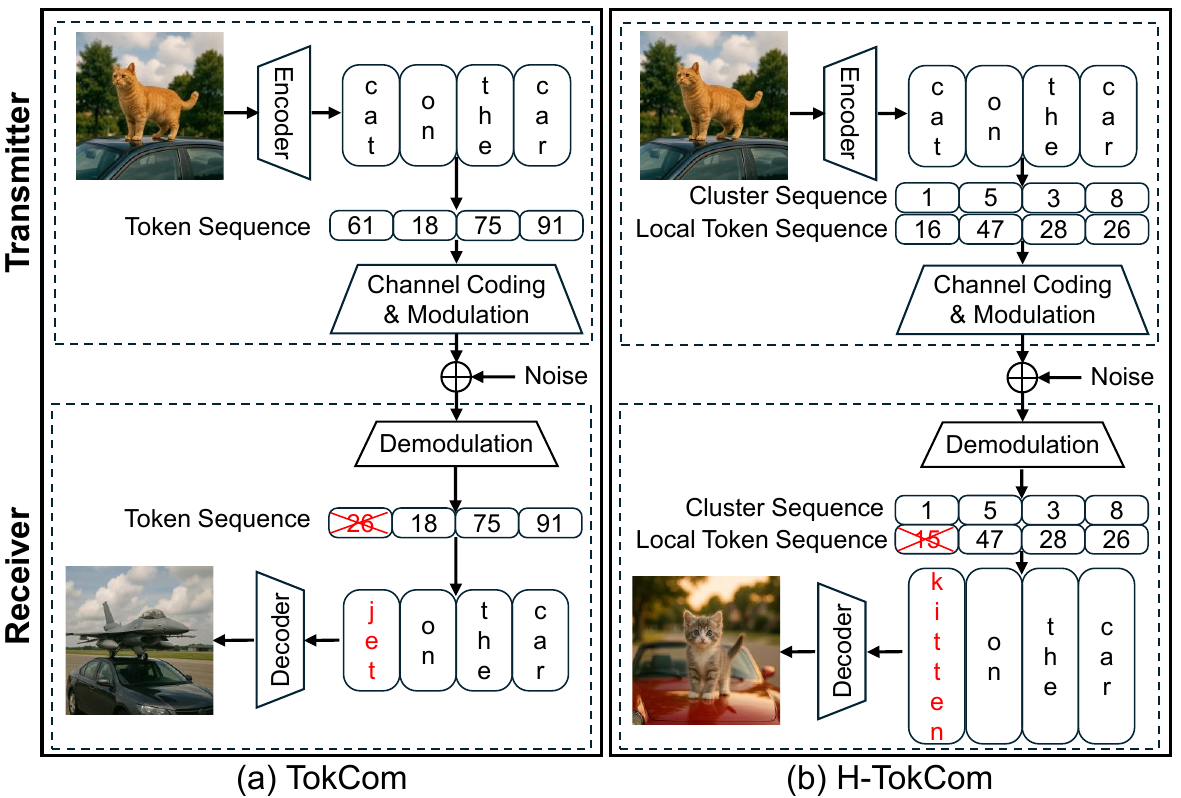}\vspace{-10pt}
  \caption{Comparison of (a) na\"{i}ve TokCom, where channel errors flip whole token indices and thus different words are decoded, and (b) the proposed hierarchical TokCom (H-TokCom), where cluster indices remain correct so only local-token indices are perturbed, yielding semantically similar words and more robust meaning over a noisy communication channel.\vspace{-20pt}}
  \label{fig:tokcom}
\end{figure}

In the context of TokCom, \emph{semantics} refer to the contextual and meaning-bearing relationships among tokens, where two tokens are semantically similar if substituting one for the other induces only minor deviation in the model’s interpretation or downstream output \cite{TokenComm_Framework2025}. 
Building on this token-level interpretation of semantics, several TokCom frameworks have recently been proposed that explicitly leverage AI-driven semantic mechanisms to improve the reliability of token delivery. For example, a text-guided TokCom scheme for image transmission is introduced in \cite{TokenComm_TextGuidedICCC2025}, where text tokens generated by a pretrained text-to-token encoder help reconstruct corrupted image tokens. {In \cite{lang_semcom}, a language-model-based processing is used to compress and robustify transmitted tokens, improving perceptual similarity under noisy channels.} Multi-modal TokCom is investigated in \cite{TokenComm_Framework2025}, where token losses in one modality are compensated by exploiting complementary information from other modalities via cross-modal attention. This line of work is extended to multiple-access scenarios in \cite{TokenDomain_MultipleAccess2025}, where token outages caused by multi-user interference are mitigated through AI-based masked token recovery. 

Collectively, these studies suggest that \emph{semantic error correction} (SEC) mechanisms have strong potential to preserve meaning more effectively than conventional channel coding techniques, such as low density parity check code, particularly from a semantic-preservation perspective \cite{SC_Short_Long}. Building on this insight, the capability of masked token recovery has also been exploited to assess the predictability of tokens and to guide higher-layer communication strategies. Specifically, semantic predictability is incorporated into packet scheduling and aggregation in \cite{TokenComm_PacketAggregation_SPAWC2025}, selective transmission and prior-aware decoding in \cite{contextaware_tokcom}, as well as retransmission control in \cite{tokcom_robust}.

Despite the demonstrated effectiveness of such AI-layer semantic recovery mechanisms, the underlying communication-layer designs in existing TokCom framework remain largely semantic-agnostic. In particular, token indices are typically converted into fixed-length bit sequences without regard to their semantic importance, and these bits are subsequently mapped to communication symbols under uniform power allocation, as illustrated in Fig. \ref{fig:tokcom}(a). As a result, consecutive symbol errors due to low \emph{signal-to-noise ratio} (SNR) can severely distort semantically important tokens, limiting the ability of AI-based recovery alone to fully compensate for communication impairments.

In this work, we propose \emph{hierarchical TokCom} (H-TokCom), a semantic-aware communication-layer design that enhances the reliability of token delivery over noisy channels by explicitly aligning semantic structure with bit-level transmission. 
This hierarchy is constructed from token-level semantics by organizing the vocabulary into semantic groups and then distinguishing individual tokens within each group.
Fig. \ref{fig:tokcom}(b) graphically illustrates an example of H-TokCom, highlighting how it effectively addresses the aforementioned challenges that remain unresolved in conventional bit-centric communication and TokCom. 
The key features of H-TokCom are summarized as follows:
\begin{itemize}[leftmargin=*]
\item \textbf{Semantic Clustering \& Bit Mapping}: The core idea of H-TokCom is to exploit semantic structure by organizing tokens into clusters. Tokens with high semantic similarity are grouped together, and each cluster is assigned a unique cluster-bit prefix. Within each cluster, individual tokens are mapped to cluster-local token bits, meaning that the same token-bit pattern may be reused across different clusters while remaining globally distinguishable through the prefixed cluster bits. Under this hierarchical representation, as long as the cluster bits are decoded correctly, any errors in the token bits cause only marginal semantic distortion.
\item \textbf{Unequal Power Allocation}: H-TokCom's reliability hinges on the accurate decoding of the cluster bits, as they specify the semantic region to which the received token belongs. Under a given transmit-power budget, we therefore prioritize the protection of cluster bits by allocating them a larger portion of power, while assigning the remaining power to the token bits. 
We construct an exponential-shaped power-mapping function that 
adaptively adjusts the cluster-bit power according to the operating SNR, 
enabling smooth and SNR-aware prioritization between semantic-region identification and fine-grained token discrimination.   
\end{itemize}
The two communication-centric designs of H-TokCom effectively preserves semantic information over noisy physical channels without relying on any AI-based module. This is validated through simulations on multiple datasets, where H-TokCom consistently outperforms baseline schemes in terms of semantic similarity across the SNR regime in which AI-based SEC method fails to operate reliably (e.g., on COCO at $\gamma=3$ dB, the semantic similarity increases from $0.206$ to $0.279$, i.e., by $0.073$ $(35.4\%)$ over Na\"{i}ve TokCom).

\vspace{-0pt}
\section{System Model}\label{sec: system model}

Consider a point-to-point TokCom, where a transmitter aims to send the semantics encapsulated in a token sequence to a receiver over noisy channels. This section describes the overall TokCom pipeline and then present our problem formulation.

\subsection{Token Preprocessing}
\subsubsection{Semantic Source to Tokens}
Let the semantic source (e.g., an image or a multimodal scene) be denoted by $\mathbf{z}$, which is first converted into a natural-language sentence $s$ and then tokenized into a sequence of discrete~tokens:
\begin{align}\label{eq: Token generation}
\mathbf{z}\rightarrow s \rightarrow (t^{(1)},t^{(2)},\cdots, t^{(N)}),
\end{align}
where $t^{(n)}$ is the $n$-th token and $N$ is the sequence~length.

Each token $t^{(n)}$ belongs to a shared vocabulary set $\mathcal{V}=\{v_1,v_2,\cdots,v_{|\mathcal{V}|}\}$, where $v_m$ denotes the $m$-th vocabulary item. Each vocabulary item $v_m$ is associated with an embedding vector $\mathbf{e}_m\in\mathbb{R}^D$, and we denote by $\mu(n)$ the vocabulary index of the $n$-th token, i.e.,
\begin{align}\label{eq:Token selection}
t^{(n)}=v_{\mu(n)}.
\end{align}
The vocabulary set $\mathcal{V}$, the tokenizer, and the embedding table $\{\mathbf{e}_m\}$ are shared between the transmitter and the receiver, ensuring compatible token interpretation at both ends.

\subsubsection{Token-to-Binary Encoding}
Each vocabulary item $v_m \in \mathcal{V}$ is mapped to an $L$-bit codeword through a mapping function $f:v_m\rightarrow \mathbf{c}_m$, namely,
\begin{align}\label{eq: mapping function}
f(v_m)=\mathbf{c}_m=(c_{m,1},c_{m,2},\cdots,c_{m,L}).
\end{align}
Using the tokenizer index $\mu(n)$ in \eqref{eq:Token selection}, the $n$-th token is converted into the following bit sequence: 
\begin{align}\label{eq: bit sequence}
\mathbf{b}^{(n)}=\mathbf{c}_{\mu(n)}
=(b_1^{(n)},b_2^{(n)},\cdots,b_L^{(n)}).
\end{align}
Accordingly, the token sequence in \eqref{eq: Token generation} is transformed into a binary representation for physical-layer transmission as
\begin{align}\label{eq: bitstream}
(t^{(1)},t^{(2)},\cdots,t^{(N)})
\rightarrow
(\mathbf{b}^{(1)},\mathbf{b}^{(2)},\cdots,\mathbf{b}^{(N)}).
\end{align}
\begin{remark}[Na\"{i}ve Bit Assignment]\label{remark: Naive Bit Assignment}
Most TokCom approaches assign binary codewords to vocabulary items without considering semantic relationships, resulting in a semantics-agnostic bit assignment. For example, as shown in Fig.\ref{fig:tokcom}(a), a transmitted token such as \textit{cat} may be decoded as an unrelated token such as \textit{jet} under a small number of bit errors, motivating the semantic-aware mapping design in the sequel.
\end{remark}

\subsection{Token Delivery over AWGN channel}

\subsubsection{Modulation} For simplicity, we consider \emph{binary phase shift keying} (BPSK) while the extension to higher-order modulation is straightforward. The $\ell$-th bit of the $n$-th token, say $b_{\ell}^{(n)}$, is mapped to real-valued symbol $x_{\ell}^{(n)}\in\mathbb{R}$ according to
\begin{align}\label{eq: symbol}
x_{\ell}^{(n)}=\sqrt{p_{\ell}^{(n)}}\,\left(2b_{\ell}^{(n)}-1\right),
\end{align}
where $p_{\ell}^{(n)}$ is the transmit power allocated to this symbol.  The resulting symbol sequence is 
\begin{align}\label{eq: symbol sequence}
({x_{1}^{(1)}, \cdots,x_{L}^{(1)}}, {x_{1}^{(2)},\cdots, x_{L}^{(2)}},\cdots, {x_{1}^{(N)},\cdots, x_{L}^{(N)}}),
\end{align} subject to a total transmit power budget $\bar{P}$,  
\begin{align}\label{eq: power constraint}
    \sum_{n=1}^{N}\ \sum_{\ell=1}^L p_{\ell}^{(n)}\ \le \bar{P}.
\end{align}

\subsubsection{Demodulation} The BPSK symbols are sequentially transmitted over an \emph{additive white Gaussian noise} (AWGN) channel yielding,
\begin{align}
    y_{\ell}^{(n)} = x_{\ell}^{(n)} + n_{\ell}^{(n)},
\end{align}
where $n_{\ell}^{(n)} \sim \mathcal{N}(0,\sigma^2)$ follows an independent and identically distributed Gaussian distribution. Under BPSK, the maximum likelihood detection reduces to
\begin{align}\label{eq: BPSK decoding}
\tilde{b}_{\ell}^{(n)} &= 
\begin{cases}
1, & y_{\ell}^{(n)} > 0,\\
0, & \text{otherwise}.
\end{cases}
\end{align}
The decoded bit sequence for the $n$-th token becomes 
$\tilde{\mathbf{b}}^{(n)}=(\tilde{b}_{1}^{(n)}, \tilde{b}_{2}^{(n)}, \cdots, \tilde{b}_{L}^{(n)})$, which is then mapped back to its corresponding token via
\begin{align}
    \tilde{t}^{(n)} ={f}^{-1}(\tilde{\mathbf{b}}^{(n)}),
\end{align}
where $f^{-1}$ is the inverse binary mapping function defined in \eqref{eq: mapping function}. Then the decoded token sequence $(\tilde{t}^{(1)},\tilde{t}^{(2)},\cdots,\tilde{t}^{(N)})$ is translated into the reconstructed sentence $\tilde{s}$.

\subsection{Semantic Similarity and Problem Formulation}

To evaluate whether the reconstructed sentence $\tilde{s}$ preserves the semantics of the original sentence $s$, we adopt a \emph{semantic similarity} metric \cite{Semantic_Sim}, defined as the cosine similarity between their sentence-level embeddings:
\begin{align}\label{eq:sim}
    \mathsf{sim}(s,\tilde{s})
= \frac{\langle \phi(s), \phi(\tilde{s}) \rangle}
       {\|\phi(s)\|\,\|\phi(\tilde{s})\|},
\end{align}
where $\phi(\cdot)$ maps a full sentence into a semantic embedding vector, e.g., using the Universal Sentence Encoder \cite{USE2018} or a BERT-based sentence encoder \cite{reimers2019sentencebert}. The operators $\langle \cdot,\cdot \rangle$ and $\|\cdot\|$ denote the inner product and Euclidean norm, respectively.

Our objective is to maximize the expected semantic similarity over sentences by jointly optimizing the token-to-bit mapping function $f(\cdot)$ in \eqref{eq: mapping function} and the power allocation under the constraint \eqref{eq: power constraint}:
\begin{align}\label{eq:opt_joint_fp}\tag{P1}
    \max_{f(\cdot), \{p_{\ell}^{(n)}\}}
     \mathbb{E}\big[\mathsf{sim}(s,\tilde{s})\big],
    \quad
    \text{s.t. \eqref{eq: power constraint}}.
\end{align}

\section{Hierarchical Token Communication}\label{sec:proposed}
This section proposes H-TokCom, a communication-layer design for efficiently solving \ref{eq:opt_joint_fp}  by leveraging the semantic relationship among vocabulary items in $\mathcal{V}$, as represented by their embedding vectors $\{\mathbf{e}_m\}$. 


\subsection{Semantic-Aware Clustering and Overview}\label{subsection: Hierarchical Bit Encoding}

This subsection presents the semantic-aware clustering that provides the basis of H-TokCom. The objective is to construct $K$ clusters, each containing at most $T$ tokens, such that tokens within the same cluster are more semantically similar in the embedding space than those in different clusters. For each of bit mapping in the following subsection, $K$ is chosen as a power of two.

Starting from singleton clusters, we iteratively merge the pair of clusters with the smallest average inter-cluster cosine distance. For two clusters $C_i$ and $C_j$, the inter-cluster distance is defined as
\vspace{-2mm}
\begin{align}\label{eq: inter-cluster distance}
D_{i,j}=\frac{1}{|C_i||C_j|} \sum_{v_u\in C_i} \sum_{v_r\in C_j} d_{u,r},
\end{align}
where
\vspace{-2mm}
\begin{align}\label{eq: distance}
d_{u,r} = 1 - \frac{\langle \mathbf{e}_u, {\mathbf{e}}_r \rangle}{\|\mathbf{e}_u\|\|\mathbf{e}_r\|}.
\end{align}

If a merged cluster violates the size limit $T$, it is partitioned by spectral bisection \cite{spectral_bisection}. This procedure continues until exactly $K$ clusters are obtained.

\begin{figure*}[t]
  \centering
  \includegraphics[width=\linewidth]{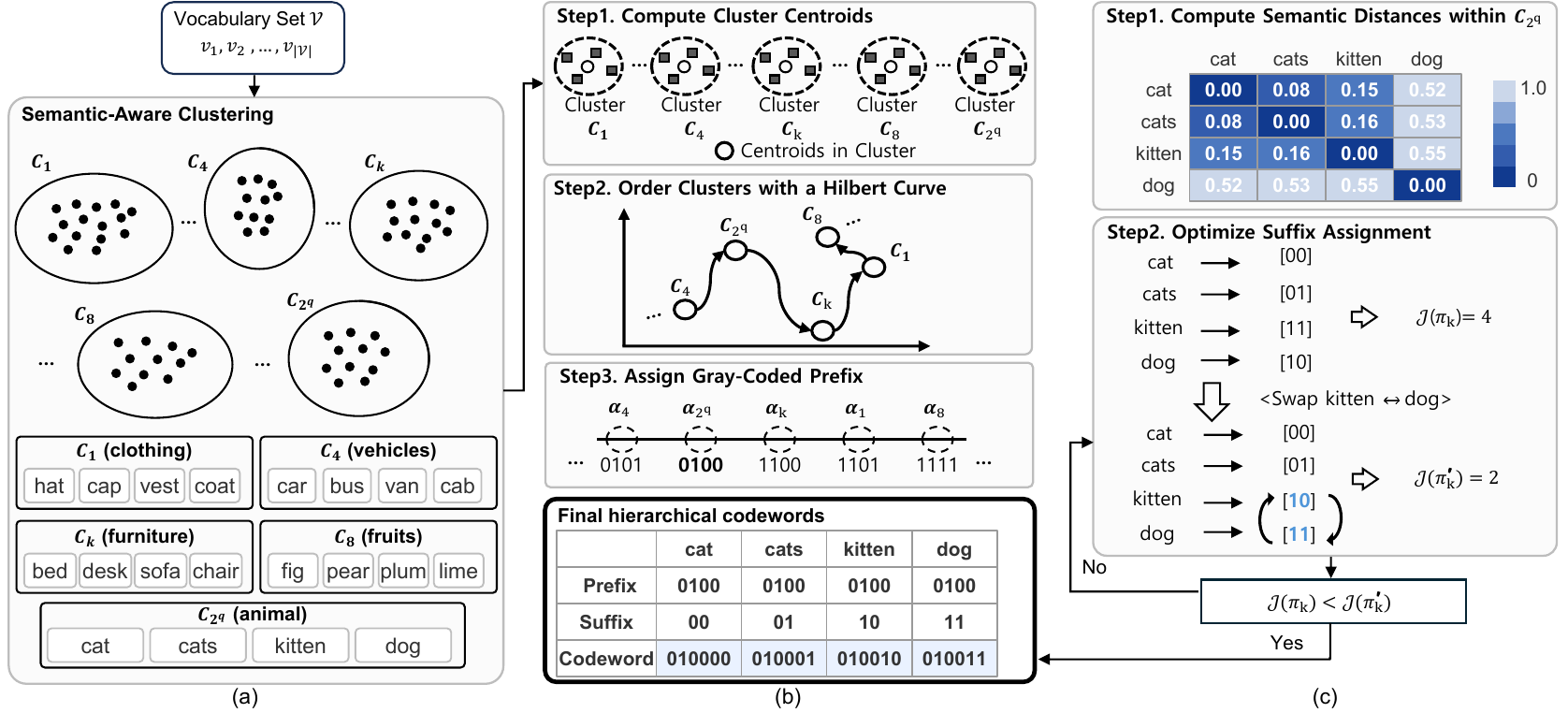}
  \vspace{-20pt}
  \caption{Illustration of the proposed hierarchical bit mapping. (a) Vocabulary items are grouped into semantic clusters. (b) Each cluster is assigned a $q$-bit prefix. (c) Tokens within each cluster are assigned $(L-q)$-bit suffixes.\vspace{-25pt}}
  \label{fig:hierarchica_bit_encoding}
\end{figure*}

Fig. \ref{fig:hierarchica_bit_encoding} (a) illustrates several clusters produced by the proposed semantic-aware clustering. Unlike the na\"{i}ve TokCom in Remark \ref{remark: Naive Bit Assignment}, this approach better preserves semantics by grouping tokens with similar meanings or categories in the same cluster. Leveraging this structure, H-TokCom incorporates two key designs. First, each vocabulary item is assigned a unique semantic-aware codeword though a hierarchical separation of  cluster-prefix and token-suffix bits, which will be explained in Sec. \ref{subsec:bit_encoding}. Second, unequal power allocation is used to provide stronger protection to the prefix bits than to the suffix bits, which will be explained in Sec. \ref{subsec:power}. 

\subsection{Hierarchical Bit Mapping}\label{subsec:bit_encoding}

We consider a $L$-bit codeword that is uniquely assigned to each vocabulary item $v_m\in\mathcal{V}$ by setting $L=\lceil\log_2|\mathcal{V}|\rceil$, where $\lceil\cdot\rceil$ is the ceil function. For hierarchical bit mapping, the codeword is divided into a $q$-bit prefix and an $(L-q)$-bit suffix, denoted by  $\boldsymbol{\alpha}_m\in\{0,1\}^q$ and $\boldsymbol{\beta}_m\in\{0,1\}^{L-q}$, respectively, as
\vspace{-3mm}
\begin{align}
\boldsymbol{\alpha}_m&=(c_{m,1}, c_{m,2},\cdots,c_{m,q}),\nonumber\\
\boldsymbol{\beta}_m&=(c_{m,q+1}, c_{m,q+2},\cdots,c_{m,L}),
\end{align}
where $q=\log_2 K$. We propose the following hierarchical bit mapping, where $\boldsymbol{\alpha}_m$ and $\boldsymbol{\beta}_m$ are constructed to represent inter- and intra-cluster semantic relationships, respectively.  

\subsubsection{Prefix-Bit Mapping}
Let $\mathbf{\zeta}_k=\frac{1}{|C_k|}\sum_{v_m\in C_k}\mathbf{e}_m$ denote the centroid of cluster $C_k$, which serves as its representative in the embedding space. To characterize inter-cluster semantic proximity, the cluster centroids are first projected onto a low-dimensional layout and then ordered via a Hilbert space-filling curve \cite{Hilbert}, thereby producing a one-dimensional representation of the centroid geometry\footnote{Compared with token-level relationships within each cluster, centroid-level relationships are coarser and thus more amenable to one-dimensional ordering.}. Let $C_{(k)}$ denote the cluster at the $k$-th position in this ordering, and let $g_q:\{0,\cdots,2^q-1\}\rightarrow\{0,1\}^q$ denotes the $q$-bit Gray-code mapping. Then, for each vocabulary item $v_m\in C_{(k)}$, the prefix bits are assigned as
\vspace{-2mm}
\begin{align}
\boldsymbol{\alpha}_m = g_q(k-1).
\label{eq:prefix_assignment}
\end{align}
By construction, all vocabulary items within the same cluster share the same prefix, while adjacent clusters are assigned prefix patterns that differ by only one bit. Accordingly, for $v_{m_1}\in C_{(k)}$ and $v_{m_2}\in C_{(u)}$, 
\begin{align}
\|\boldsymbol{\alpha}_{m_1}\oplus\boldsymbol{\alpha}_{m_2}\|_1=
\begin{cases}
0, & k=u,\\
1,  & |k-u|=1,
\end{cases}
\end{align}
where $\oplus$ is the bitwise XOR and $\|\cdot\|_1$ represents the $\ell_1$-norm. 
This procedure is illustrated in  Fig. \ref{fig:hierarchica_bit_encoding}(b).

\subsubsection{Suffix-Bit Mapping}

Token-level semantic relationships within each cluster are more fine-grained than centroid-level proximity in the embedding space and therefore not adequately captured by the one-dimensional ordering described above. Instead, they should be represented in the $(L-q)$-bit sequence space, where proximity is measured by Hamming distance. To align the semantic similarity with this bit-level structure, we define the target Hamming distance between two items $v_u,v_r\in C_k$ as
\vspace{-1mm}
\begin{align}\label{eq: target Hamming Dist}
\bar{D}_{u,r}=(L-q)\, d_{u,r}^{\rho},
\end{align}
where $d_{u,r}$ is the semantic distance in \eqref{eq: distance}, and $\rho>0$ controls the nonlinear mapping from semantic distance to target Hamming distance. 

Let $\pi_k:C_k\rightarrow \{0,1\}^{L-q}$ denote a one-to-one suffix-bit assignment for the vocabulary items in cluster $C_k$. The resultant distortion is defined as
\vspace{-0.5mm}
\begin{align}\label{eq: distortion}
\mathcal{J}(\pi_k)
=
\sum_{\forall v_u,v_r\in C_k}
\Big(\|\pi_k(v_u)\oplus\pi_k(v_r)\|_1-\bar{D}_{u,r}\Big)^2,
\end{align}
which quantifies the mismatch between the assigned and target Hamming distances. The suffix bits are then determined by the optimal assignment 
\vspace{-0.5mm}
\begin{align}
\pi_k^*=\arg\min_{\pi_k}\mathcal{J}(\pi_k),
\qquad
\boldsymbol{\beta}_m=\pi_k^*(v_m), \ \forall v_m\in C_k.
\end{align}
Since exhaustive search over all feasible assignments is combinatorial, we employ a low-complexity iterative swapping procedure. Starting from an initial assignment, the suffix-bit assignment between two vocabulary items are swapped whose swap produces the largest decrease in $\mathcal{J}(\pi_k)$. The procedure terminates when no further improving swap exists. This procedure is illustrated in  Fig. \ref{fig:hierarchica_bit_encoding}(c).
\begin{remark}[Computation Complexity] There are $\binom{|C_k|}{2}=O(|C_k|^2)$ candidate swaps at each iteration. Since the effect of swapping two items can be evaluated by updating only the $O(|C_k|)$ affected pairwise terms, the per-iteration complexity is $O(|C_k|^3)$. Accordingly, if the algorithm runs for $I$ iterations, the overall complexity is $O(I|C_k|^3)$.
\end{remark}

The resulting hierarchical mapping is defined as
\vspace{-2mm}
\begin{align}\label{eq:hierarchical_bit_mapping_function}
\mathbf{c}_m=f_{\mathrm{hier}}(v_m)=\boldsymbol{\alpha}_m\parallel\boldsymbol{\beta}_m,
\end{align}
where $\parallel$ denotes concatenation. Thus, for the $n$-th token in a sentence, the transmitted bit sequence can be decomposed into its prefix and suffix parts as
\vspace{-2mm}
\begin{align}\label{eq:prefix_suffix_for_token}
\boldsymbol{\Lambda}^{(n)}=\boldsymbol{\alpha}_{\mu(n)}, \quad
\boldsymbol{\lambda}^{(n)}=\boldsymbol{\beta}_{\mu(n)},
\end{align}
where the $n$-th token's vocabulary\! index $\!\mu(n)\!$ is defined in  \eqref{eq:Token selection}. 

\subsection{Cluster-Prioritized Power Allocation}\label{subsec:power}

Given the hierarchical bit mapping, we adopt a cluster-prioritized power allocation strategy by allocating more power to the cluster-prefix bits than to the token-suffix bits, since reliable delivery of the prefix bits $\boldsymbol{\Lambda}^{(n)}$ is critical for preserving the semantic region of each token. 

Under the total power budget $\bar{P}$ in \eqref{eq: power constraint}, we allocate the same per-token power budget to all tokens:
\begin{align}
\sum_{\ell=1}^L p_{\ell}^{(n)} \le \frac{\bar{P}}{N}\triangleq P_{\mathrm{tok}}.
\end{align}
Besides, we consider uniform power allocation within the prefix and suffix groups. Under the per-token budget $P_{\mathrm{tok}}$, the prefix-symbol power lies between the equal-power case and the prefix-only allocation case:
\begin{align} \label{eq: power range}
\frac{P_{\mathrm{tok}}}{L} \leq p_{\ell}^{(n)} \leq \frac{P_{\mathrm{tok}}}{q}, \quad \ell=1,\cdots, q.
\end{align}

Let $\varepsilon$ denote the target \emph{symbol error rate} (SER) for the prefix BPSK symbols. Its feasible range is given by 
\begin{align}\label{eq: lower and upper}
\varepsilon_{\textrm{lower}}=Q_e\left(\frac{P_{\mathrm{tok}}}{q\sigma^2}\right),\
\varepsilon_{\textrm{upper}}=Q_e\left(\frac{P_{\mathrm{tok}}}{L\sigma^2}\right),
\end{align}
where $Q_e(x)=\frac{1}{\sqrt{\pi}}\int_{\sqrt{x}}^\infty e^{-u^2} du$ represents the SER when the symbol SNR is $x$. 

\begin{figure}[t]
    \centering
    \includegraphics[width=\linewidth]{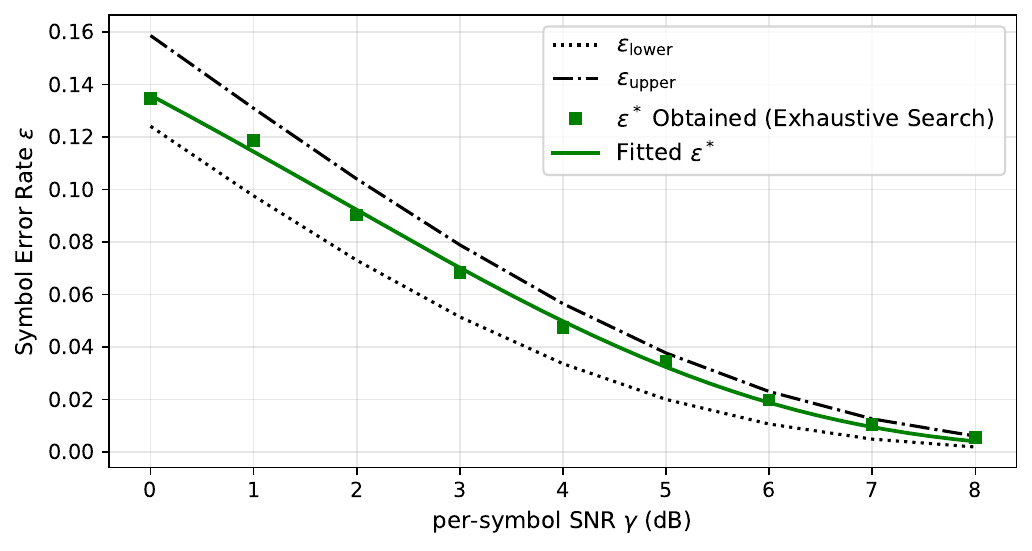}\vspace{-10pt}\caption{The optimal target SER $\varepsilon^*$ in \eqref{eq: optimal power allocation} as a function of the per-symbol SNR $\gamma\triangleq\frac{P_{\mathrm{tok}}}{L\sigma^2}$, obtained through fitting exhaustive-search results. For comparison, we plot its design bounds $\varepsilon_{\textrm{lower}}$ and $\varepsilon_{\textrm{upper}}$ defined in \eqref{eq: lower and upper}. \vspace{-20pt}}
    \label{fig:SER_target}
\end{figure}

Within the feasible range, the target SER  $\varepsilon$ directly determines the power allocation to both prefix and suffix symbols:
\begin{align}
p_{\ell}^{(n)}=
\begin{cases}
\sigma^2 Q_e^{-1}(\varepsilon), & \ell\in\{1,\cdots,q\},\\
\frac{P_{\mathrm{tok}}-q\sigma^2 Q_e^{-1}(\varepsilon)}{L-q}, & \ell\in\{q+1,\cdots,L\}.
\end{cases}
\end{align}
Thus, the entire power allocation strategy is controlled by the single parameter $\varepsilon$. Its operating value is determined as a function of the per-symbol SNR $\gamma \triangleq\frac{P_{\mathrm{tok}}}{L\sigma^2}$ through exhaustive search and curve fitting:
\begin{align}\label{eq: optimal power allocation}
\varepsilon^*(\gamma)=\eta e^{-\theta \gamma},
\end{align}
where $\eta=0.1888$ and $\theta=0.5740$. 
Fig.~\ref{fig:SER_target} compares the fitted curve (solid green line) with the exhaustive-search results (green square markers), showing that the exponential approximation closely matches the optimal $\varepsilon^*$ over the SNR range of interest.

Relative to the design bounds in \eqref{eq: lower and upper}, the mapping $\varepsilon^*(\gamma)$ exhibits three main behaviors. First, it decreases monotonically with $\gamma$, so the target SER for the cluster-prefix symbols becomes stricter as the overall SNR increases. Second, in the low-SNR regime, $\varepsilon^*(\gamma)$ stays below $\varepsilon_{\mathrm{upper}}$ and well above $\varepsilon_{\mathrm{lower}}$, indicating a moderate bias toward prefix protection without approaching prefix-only allocation. Third, as $\gamma$ increases, $\varepsilon^*(\gamma)$ moves closer to $\varepsilon_{\mathrm{upper}}$, implying convergence to the equal-power operating point at high SNR.

\subsection{Semantic-Aware Token Reconstruction}\label{subsection: Token Reconstruction}

The received BPSK symbols are decoded according to \eqref{eq: BPSK decoding}, yielding the decoded prefix and suffix bits $\tilde{\boldsymbol{\Lambda}}^{(n)}$ and $\tilde{\boldsymbol{\lambda}}^{(n)}$ for the $n$-th token. First, the decoded prefix $\tilde{\boldsymbol{\Lambda}}^{(n)}$ identifies the corresponding cluster, denoted by $C_{k(n)}$. Second, the decoded suffix $\tilde{\boldsymbol{\lambda}}^{(n)}$ is matched within the set $\{\boldsymbol\beta_m\}_{v_m\in C_{k(n)}}$, whose suffix has the minimum Hamming distance to $\tilde{\boldsymbol{\lambda}}^{(n)}$, namely, 
\begin{align}\label{eq: finding the token suffix bits}
m^*(n)= \arg\min_{\{m\,|\,v_m\in C_{k(n)}\}}
\big\|
\tilde{\boldsymbol{\lambda}}^{(n)} \oplus \boldsymbol{\beta}_m
\big\|_1.
\end{align}
The reconstructed token is then given by
\begin{align}
\tilde{t}^{(n)} = f_{\mathrm{hier}}^{-1}\!\left(\tilde{\boldsymbol{\Lambda}}^{(n)} \parallel \boldsymbol{\beta}_{m^*(n)}\right),
\end{align}
where $f_{\mathrm{hier}}^{-1}(\cdot)$ is the inverse of the function $f_{\mathrm{hier}}(\cdot)$ in \eqref{eq:hierarchical_bit_mapping_function}. 

\section{Simulation Results}\label{Sec: Simulation}

This section evaluates and compares the following schemes:
\begin{itemize}
\item \textbf{Na\"{i}ve TokCom:}
    Each token is represented by a fixed-length $L$-bit sequence without semantic clustering, and all $L$ bits are transmitted with uniform power under the per-token budget $P_{\mathrm{tok}}$.
\item \textbf{SEC:}
    We consider a conventional SEC \cite{contextaware_tokcom} on top of Na\"{i}ve TokCom, where token candidates are generated from contextual predictions of a masked language model. The detailed setting is omitted due to the page limit. 
\item \textbf{H-TokCom without Power Control:}
    The hierarchical bit mapping in Sec.~\ref{subsec:bit_encoding} is used, but each bit's transmit power is equally allocated under the same budget $P_{\mathrm{tok}}$.

\item \textbf{H-TokCom (Proposed):}
    The proposed H-TokCom applies both the hierarchical bit mapping and the cluster-prioritized power allocation under the same budget $P_{\mathrm{tok}}$, all of which are explained in Sec.~\ref{sec:proposed}.

\end{itemize}


\subsection{Simulation Setting}

\subsubsection{Experimental Setup}
We evaluate all schemes as a function of per-symbol SNR $\gamma$, defined by normalizing the per-token power budget $P_{\mathrm{tok}}$ by the product of the code length $L$ and noise variance $\sigma^2$. We set $P_{\mathrm{tok}}=3$ and $L=16$, while varying per-symbol SNR $\gamma$ from $0$ to $8.0$ $\mathrm{dB}$. Each experiment is repeated $100$ times to average over independent channel realizations. 

Next, we evaluate each scheme on $30$ natural-language sentences randomly sampled from each of three datasets: COCO Captions \cite{MSCOCO2015}, QQP \cite{QQP_Sharma2019}, and Flickr30k \cite{flickr30k}. Sentences are tokenized using the pretrained CLIP tokenizer whose vocabulary size is $|\mathcal{V}|= 49{,}408$. For H-TokCom, each $L$-bit codeword is decomposed into $q=12$ cluster-prefix bits  and $L-q=4$ token-suffix bits, yielding $K=2^{q}=4096$ semantic clusters, each with maximum size $T=2^{L-q}=16$. We use the pretrained all-MiniLM-L6-v2 sentence encoder \cite{allMiniLML6v2} to generate semantic embeddings $\phi(s)$. 

\begin{figure*}[t]
  \centering
  \includegraphics[width=\textwidth]{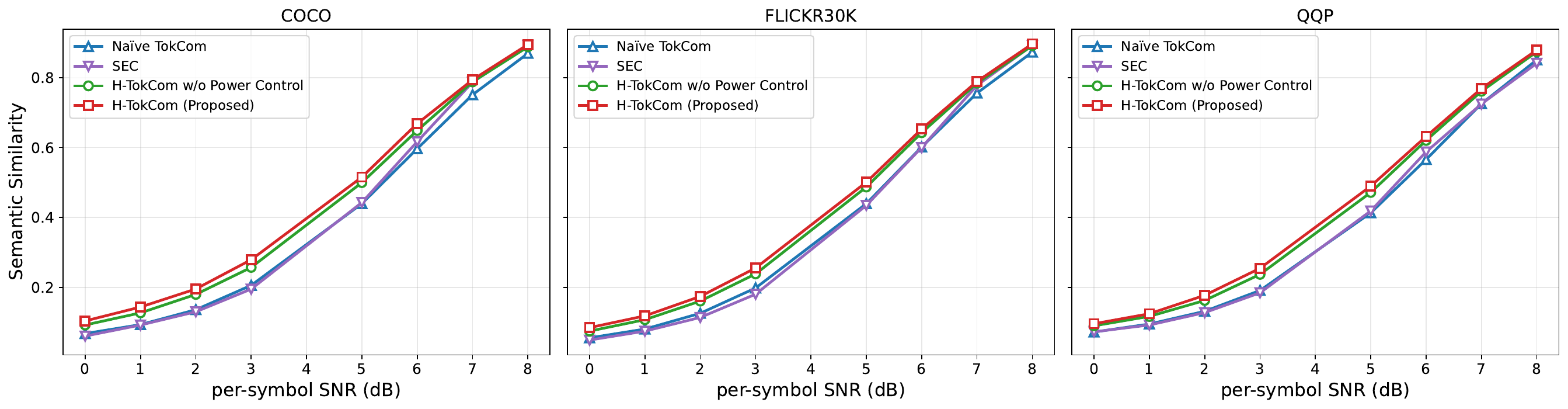}
  \vspace{-20pt}\caption{Average semantic similarity versus per-symbol SNR $\gamma$ on three datasets: COCO, Flickr30k, and QQP. In each plot, we compare four schemes: Na\"{i}ve TokCom, H-TokCom without power control, H-TokCom, and Na\"{i}ve TokCom with semantic error correction (SEC).\vspace{-10pt}}
  \label{fig:Sim_Result_Four}
\end{figure*}
\subsection{Semantic Similarity}



Fig.~\ref{fig:Sim_Result_Four} shows the average semantic similarity versus the per-symbol SNR $\gamma$ on COCO, Flickr30k, and QQP for the four schemes under consideration. A consistent trend is observed across all three datasets. First, both hierarchical schemes, namely H-TokCom without Power Control and H-TokCom, consistently outperform Na\"{i}ve TokCom over the entire SNR range. This confirms that separating cluster-prefix information from intra-cluster token information is effective for preserving token semantics over noisy channels. Even without unequal power allocation, the hierarchical bit structure itself already improves semantic robustness by reducing the chance that a small number of bit errors leads to a semantically unrelated token. For example, at $\gamma=3$ dB, H-TokCom improves the semantic similarity over Na\"{i}ve TokCom from $0.206$ to $0.279$ on COCO $(+0.073, 35.6\%)$, from $0.198$ to $0.256$ on Flickr30k $(+0.057, 28.8\%)$, and from $0.191$ to $0.255$ on QQP $(+0.063, 33.1\%)$.

Second, H-TokCom consistently outperforms H-TokCom without Power Control at every SNR point. This shows that cluster-prioritized power allocation provides an additional gain beyond the hierarchical representation itself. The gain is particularly pronounced in the low-to-moderate SNR regime, where protecting the prefix bits helps preserve the coarse semantic region of each token. As the SNR increases, the gap becomes smaller, since the prefix bits can already be decoded more reliably even without unequal power allocation.

Unlike the hierarchical schemes, SEC exhibits a different behavior. In the low-SNR regime, SEC provides only limited gains over Na\"{i}ve TokCom and may even degrade performance. This indicates that when the received token sequence is heavily corrupted, receiver-side correction based mainly on contextual plausibility cannot reliably recover the original semantics and may instead generate linguistically natural but semantically mismatched outputs. As the SNR increases, SEC becomes more competitive because the received sequence becomes sufficiently reliable for language-model-based refinement. Nevertheless, the proposed H-TokCom schemes still achieve the strongest overall performance across the entire SNR range.


\section{Conclusion}
\vspace{-5pt}
In this paper, we have developed H-TokCom, a novel TokCom framework incorporating a two-tier semantic hierarchy into communication-centric design. Its key features are  hierarchical bit mapping that separates cluster-prefix bits and token-suffix bits to better preserve semantic relationships and unequal power allocation that prioritizes the more semantically informative cluster-prefix bits. Simulation results show that H-TokCom outperforms conventional TokCom benchmarks that do not exploit semantic structure. As future work, we will develop a hierarchy-aware SEC technique to further integrate communication-centric and AI-based approaches.


\vspace{-5pt}
\bibliographystyle{IEEEtran} 
\vspace{-5pt}
\bibliography{ref}    
\end{document}